\documentclass{PoS}

\title{
Systematics analyses on nucleon isovector observables in 2+1-flavor dynamical domain-wall lattice QCD near physical mass
}

\ShortTitle{DWF nucleon \(g_A\) systematics}

\author{
\speaker{
Shigemi Ohta}  
for RBC and UKQCD Collaborations\\
        Institute of Particle and Nuclear Studies, KEK, Tsukuba, Ibaraki 3050801, Japan\\
        Department of Particle and Nuclear Physics, SOKENDAI, Hayama, Kanagawa, 2400193, Japan\\
        RIKEN BNL Research Center, BNL, Upton, NY 11973, USA\\
        E-mail: \email{shigemi.ohta@kek.jp}}

\abstract{
Analyses on possible systematics in some isovector nucleon observables in the RBC+UKQCD 2+1-flavor dynamical domain-wall fermion (DWF) lattice-QCD are presented.
The vector charge, axial charge, quark momentum and helicity fractions, and transversity are discussed using mainly the Iwasaki\(\times\)DSDR ensemble at pion mass of 170 MeV.
No autocorrelation issue is observed in the vector charge and quark momentum and helicity fractions. Blocked Jack-knife analyses expose significant growth of estimated error for the axial charge with increasing block sizes that are similar to or larger than the known autocorrelation time of the gauge-field topological charge.
Similar growth is seen in the transversity.
These two observables, however, do not seem correlated with the topological charge.

\vspace{-177mm}\parbox{\textwidth}{\flushright\large\rm \hfill KEK-TH-1770, RBRC-1097}\vspace{174mm}
}

\FullConference{The 32nd International Symposium on Lattice Field Theory,\\
		23-28 June, 2014\\
		Columbia University New York, NY}

\begin{document}

\section{Introduction}

\begin{figure}[b]
\includegraphics[width=0.48\textwidth]{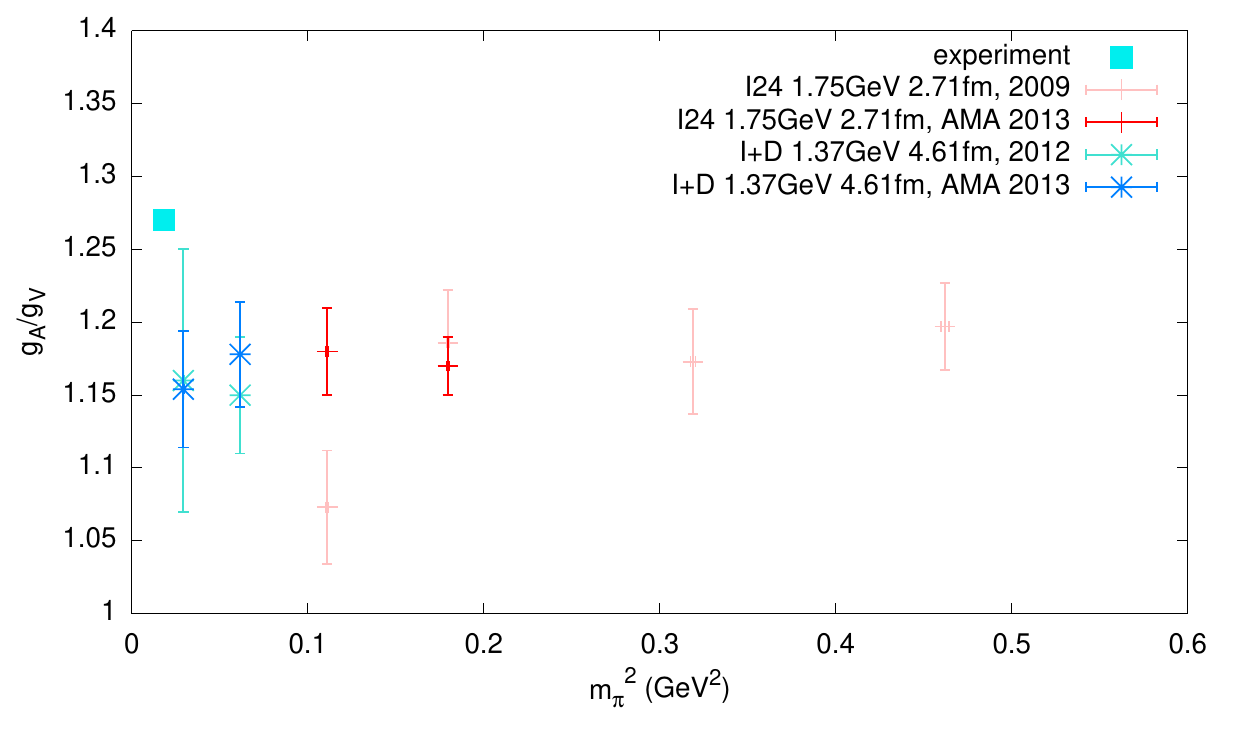}
\includegraphics[width=0.48\textwidth]{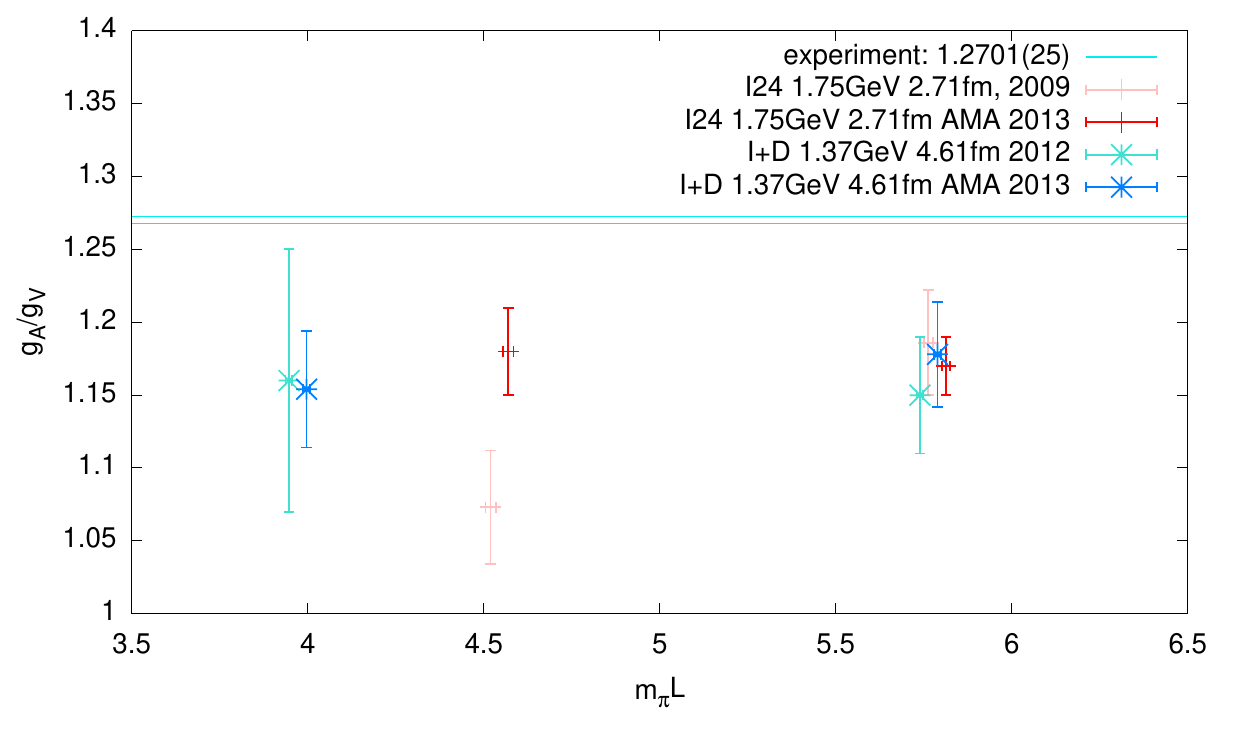}
\caption{\label{fig:AVmpi}
Dependence of the ratio, \(g_A/g_V\), of isovector axial charge, \(g_A\), and vector charge, \(g_V\), calculated with recent RBC+UKQCD 2+1-flavor dynamical DWF ensembles, on the pion mass squared, , \(m_\pi^2\) (left), and the finite-size scaling parameter, \(m_\pi L\) (right).
Solid symbols are the present results, while the faded ones are from our earlier publications.
While the precision agreement of the two ensembles at \(m_\pi L\) of about 5.8 has increased its significance (right),
the dependence on \(m_\pi^2\) does not show any sign of approaching the experiment as the pion mass squared decreases toward physical value (left).
The experimental value quoted here is 1.2701(25) from PDG 2013 rather than 1.2723(23) from  the latest PDG 2014.
}
\end{figure}

The RIKEN-BNL-Columbia (RBC) Collaboration have been investigating nucleon structure using 2+1-flavor dynamical domain-wall-fermions (DWF) lattice-QCD ensembles generated jointly with the UKQCD Collaboration.
The observables we are looking at are isovector vector- and axialvector-current form factors and some low moments of isovector structure functions, namely the quark momentum fraction, \(\langle x \rangle_{u-d}\), helicity fraction, \(\langle x \rangle_{\Delta u - \Delta d}\), transversity, \(\langle 1\rangle_{\delta u - \delta d}\) ,and twist-3, \(d_1^{u-d}\), moments.
None of them, except the trivial case of conserved vector charge, \(g_V\), agrees with experiment.
However the calculations so far have been done with the degenerate up- and down-quark mass set higher than its physical value, and the majority of the observables seem to trend to experiment as the mass is set lighter \cite{Lin:2014saa}.
An important exception to the trend is the axial charge, \(g_A\), or its ratio to the vector charge, \(g_A/g_V\):
As was reported in Lattice 2013 \cite{Ohta:2013qda} (see Fig.\ \ref{fig:AVmpi}),
the calculated values of the ratio, \(g_A/g_V\), underestimates the experimental value of 1.2723(23) \cite{PDG:2014} by about 10 \% and do not depend much on the pion mass, \(m_\pi\), in the range from about 420 MeV down to 170 MeV from four recent RBC+UKQCD 2+1-flavor dynamical DWF ensembles \cite{Allton:2008pn,Aoki:2010dy,Arthur:2012opa,I4864:2014}.
Worse, the central values appear to be moving away from the experiment as the mass is set lighter, though all values are consistent with each other within statistical errors that do not cover the experiment.
As was also reported last year, we observed unusually long-range autocorrelation in the two observables (see Fig.\ \ref{fig:gAgV2halves}.)
Here I report our exploration for the cause of this systematics.

\section{Numerics}

The four lattice ensembles are described in our earlier publications \cite{Allton:2008pn,Aoki:2010dy,Arthur:2012opa,I4864:2014}: the two heavier ensembles were generated with Iwasaki gauge action at \(\beta=2.13\), corresponding to an estimated inverse lattice spacing of \(a^{-1}=1.729(4)\) GeV, and pion mass values of about 420 and 330 MeV, and the two lighter ones were with Iwasaki\(\times\)DSDR (dislocation suppressing determinant ratio) gauge action with \(\beta=1.75\) or \(a^{-1}=1.371(10)\) GeV and pion mass of about 250 and 170 MeV.
We use conventional 3-point to 2-point correlation function ratios to calculate our isovector nucleon observables \cite{Lin:2008uz,Yamazaki:2008py,Yamazaki:2009zq,Aoki:2010xg}.
The DWF scheme provides major advantage with its continuum-like chiral and flavor symmetries.
They make non-perturbative renormalizations of relevant currents straight forward.
In particular the isovector vector and axialvector local currents share common renormalization, \(Z_A = Z_V\), up to small \(O(a^2)\) discretization error.
We optimized our Gaussian-smeared nucleon source and sink, as well as source-sink separation, and confirmed that excited-state contamination is negligible \cite{Ohta:2013qda}.
We use AMA technology \cite{Blum:2012uh} to enhance our statistics for the 330-MeV and 170-MeV ensembles, and increased statistics by simply analyzing more configurations for the 420-MeV and 250-MeV ensembles \cite{Lin:2014saa,Ohta:2013qda}.

\section{Axial charge as of Lattice 2013}

\begin{figure}[b]
\includegraphics[width=0.48\textwidth]{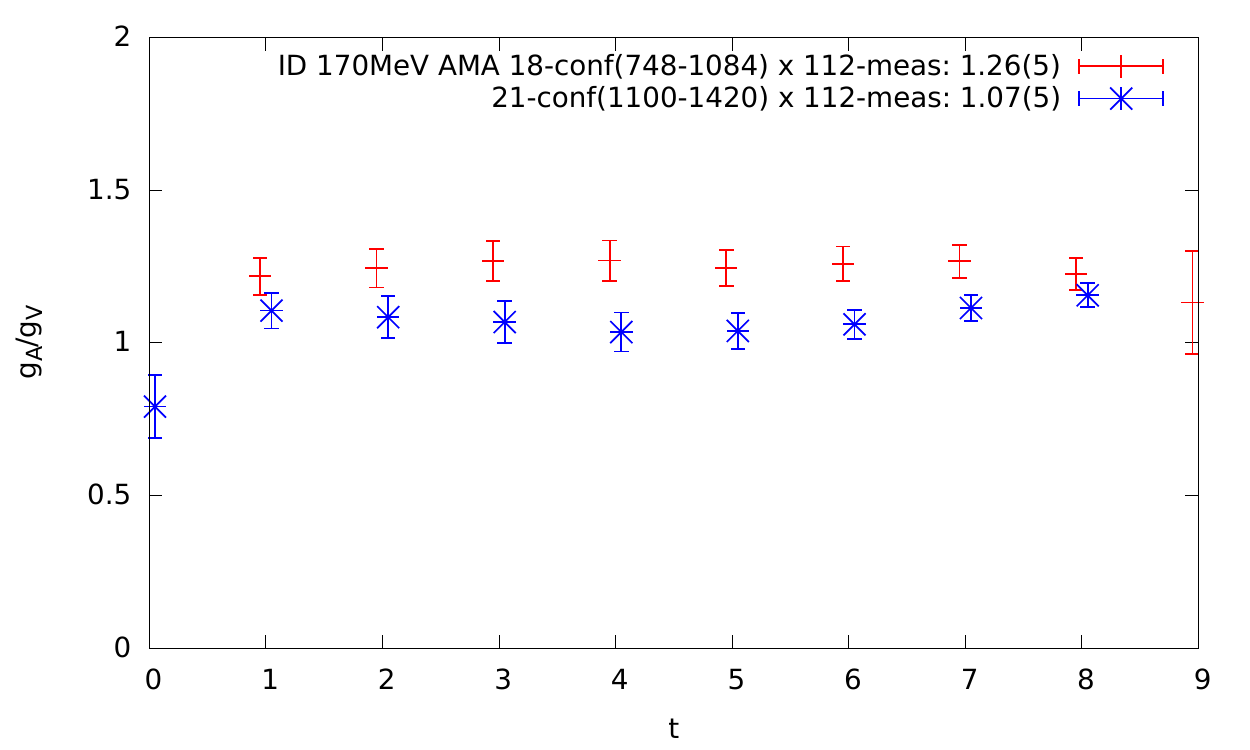}
\includegraphics[width=0.48\textwidth]{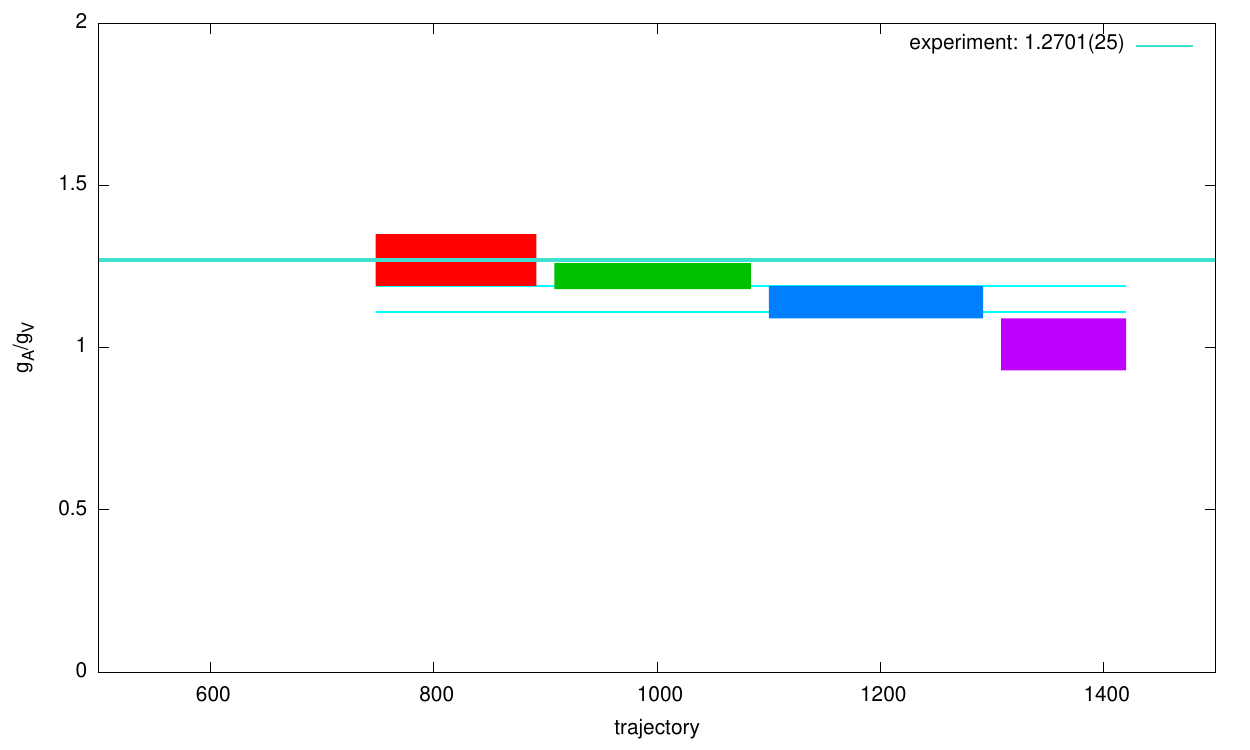}
\caption{\label{fig:gAgV2halves}
On the left we present the plateaux of the ratio, \(g_A/g_V\), for the first (trajectory from 748 to 1084, red) and the second (1100 to 1420, blue) halves, respectively: fitted in the range from 2 to 7 lattice units, the values of 1.26(5) for the first and 1.07(5) for the second are almost four standard deviations away from each other.
On the right we present quarter-wise average along the hybrid MD time, from 748 to 892, 908 to 1084, 1100 to 1292, and 1308 to 1420: the values seem to drift monotonically from what is consistent with the experiment of 1.2701(25) in the first quarter to a value around 1.0 in the last quarter.
}
\end{figure}
Taking advantage of the good chiral and flavor symmetries provided by the DWF scheme, we were the first \cite{Yamazaki:2008py} to identify a deficit in the ratio, \(g_A/g_V\), of isovector axial, \(g_A\), and vector, \(g_V\), charges, of about 10 \%.
By comparing our results with earlier calculations which also had seen similar effect but without noticing, we had conjectured the deficit is related to the relatively small lattice volumes \cite{Yamazaki:2008py}:
there seems a region in \(m_\pi L\), the dimensionless product of the pion mass \(m_\pi\) and the lattice linear extent \(L\), where the calculated value of the ratio seems to scale with it \cite{Yamazaki:2008py}.
The majority of calculations that have emerged since then align well with this trend with few exceptions \cite{Martha:LAT2014,Horsley:2013ayv}.

Then Mainz group raised an interesting question if contamination from excited nucleons were properly screened out in these calculations \cite{Capitani:2012gj}.
Though we had reasons to think our own earlier calculations had properly screened out such contaminations, as described in our earlier publications, to firmly exclude this possibility it was desirable to enhance our statistics.
Fortunately with the development of the AMA technology \cite{Blum:2012uh} combined with increased availability of modern computers, we have enhanced our statistics: in my talk at Lattice 2013 last year I reported our calculations do not suffer from such excited-state contaminations \cite{Ohta:2013qda}.
Using those values that are free of excited-state contaminations, I also presented summary figures
for the ratio, \(g_A/g_V\), in Fig.\ \ref{fig:AVmpi}.
Here the solid symbols reflect the numbers from improved statistics last year, and in particular with AMA for the two ensembles with smaller \(m_\pi L\) (or \(m_\pi = 170\) MeV and 330 MeV).

We are obviously suffering from some systematics that make our calculations undershoot the experimental value of \(g_A/g_V = 1.2723(23)\) \cite{PDG:2014}.
Indeed I also reported last year possible signs of inefficient sampling as summarized in Fig.\ \ref{fig:gAgV2halves}.
Also reported were a) no such under sampling is seen in any other isovector observables that had been looked at, the vector charge, \(g_V\), quark momentum fraction, \(\langle x \rangle_{u-d}\) and quark helicity fraction, \(\langle x \rangle_{\Delta u-\Delta d}\), and b) blocked-jackknife analyses with block size of 2 showed strong correlation of two successive gauge configurations for \(g_A\) and \(g_A/g_V\) but not for the other observables.

\section{Exploration as of Lattice 2014}

\begin{figure}[b]
\includegraphics[width=0.48\textwidth]{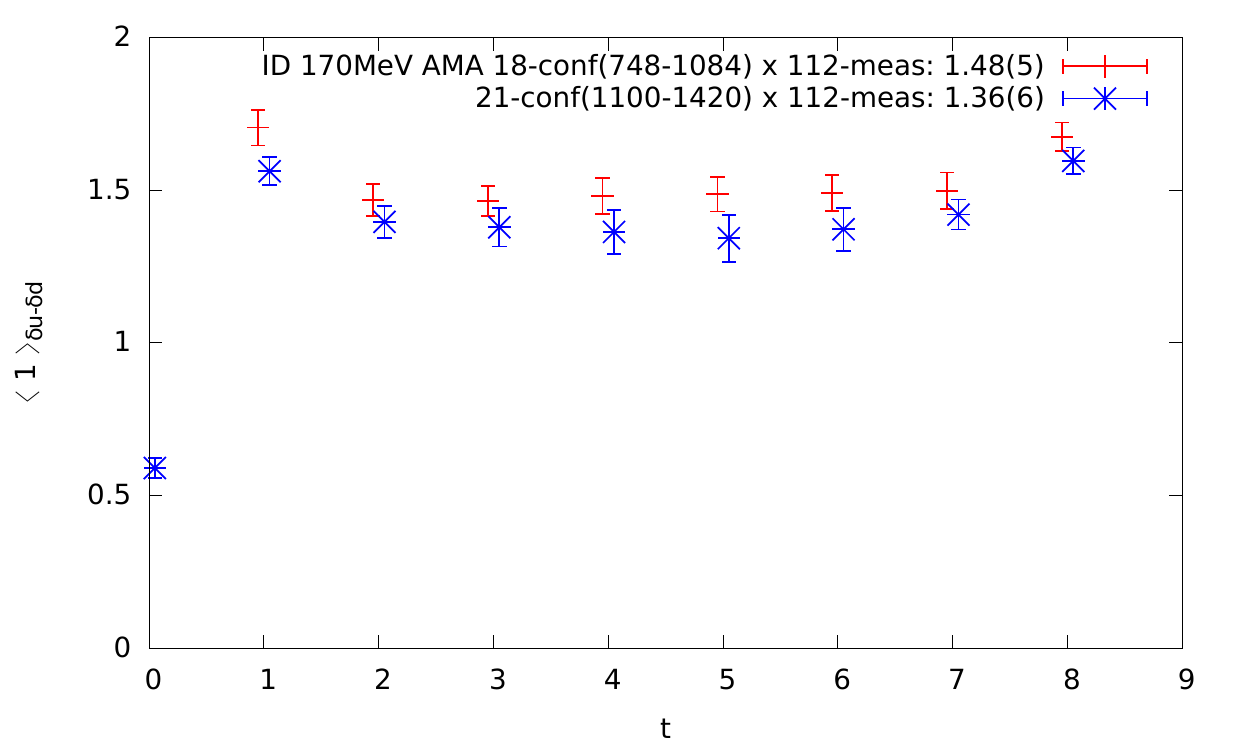}
\includegraphics[width=0.48\textwidth]{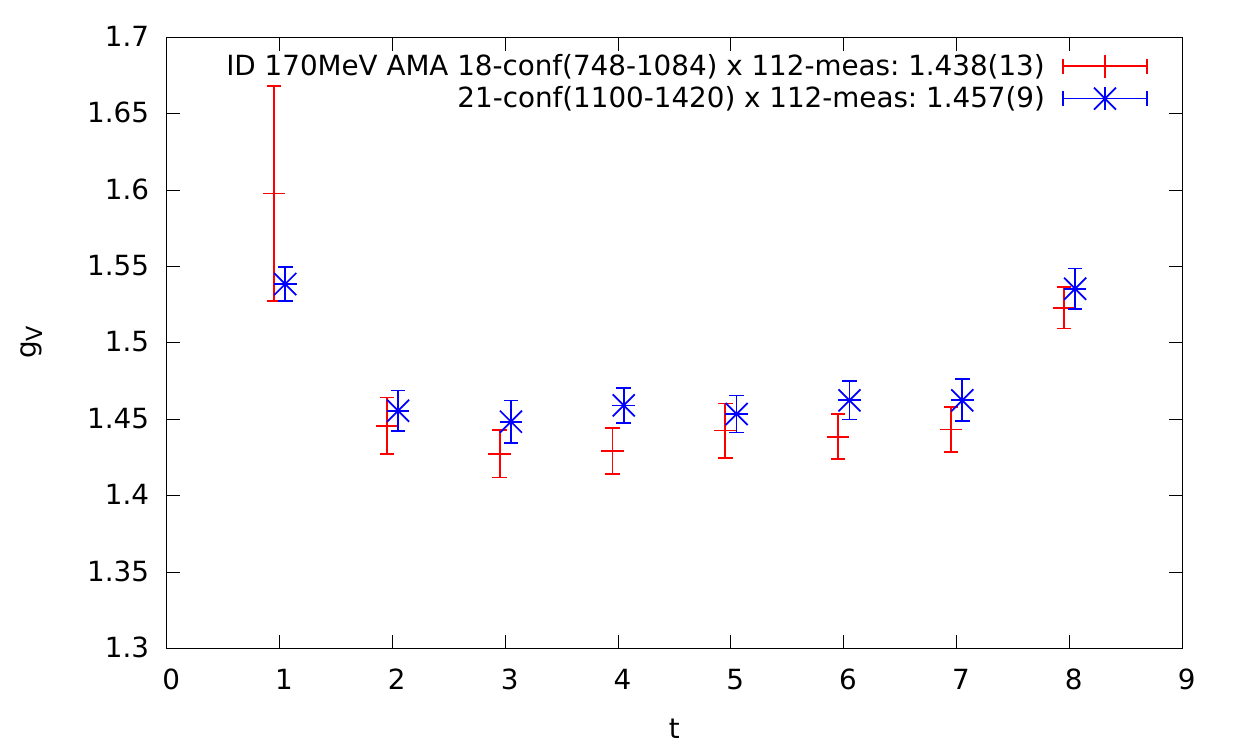}\\
\includegraphics[width=0.48\textwidth]{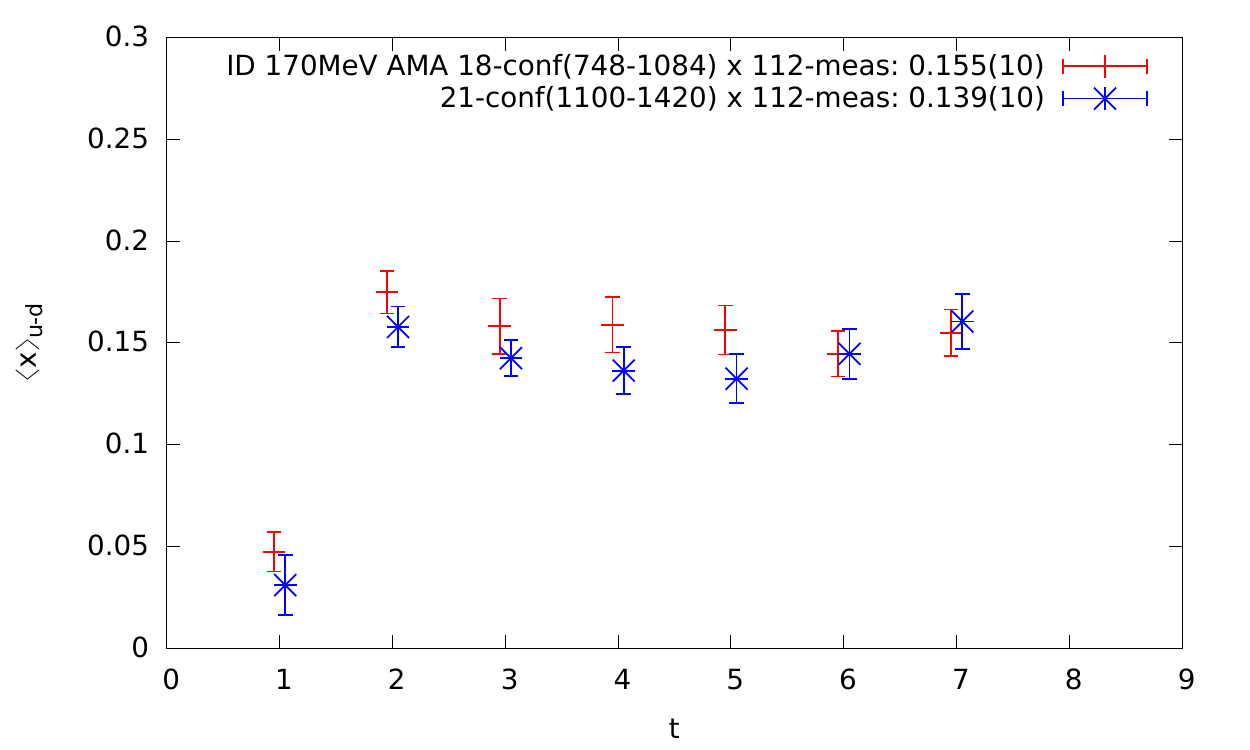}
\includegraphics[width=0.48\textwidth]{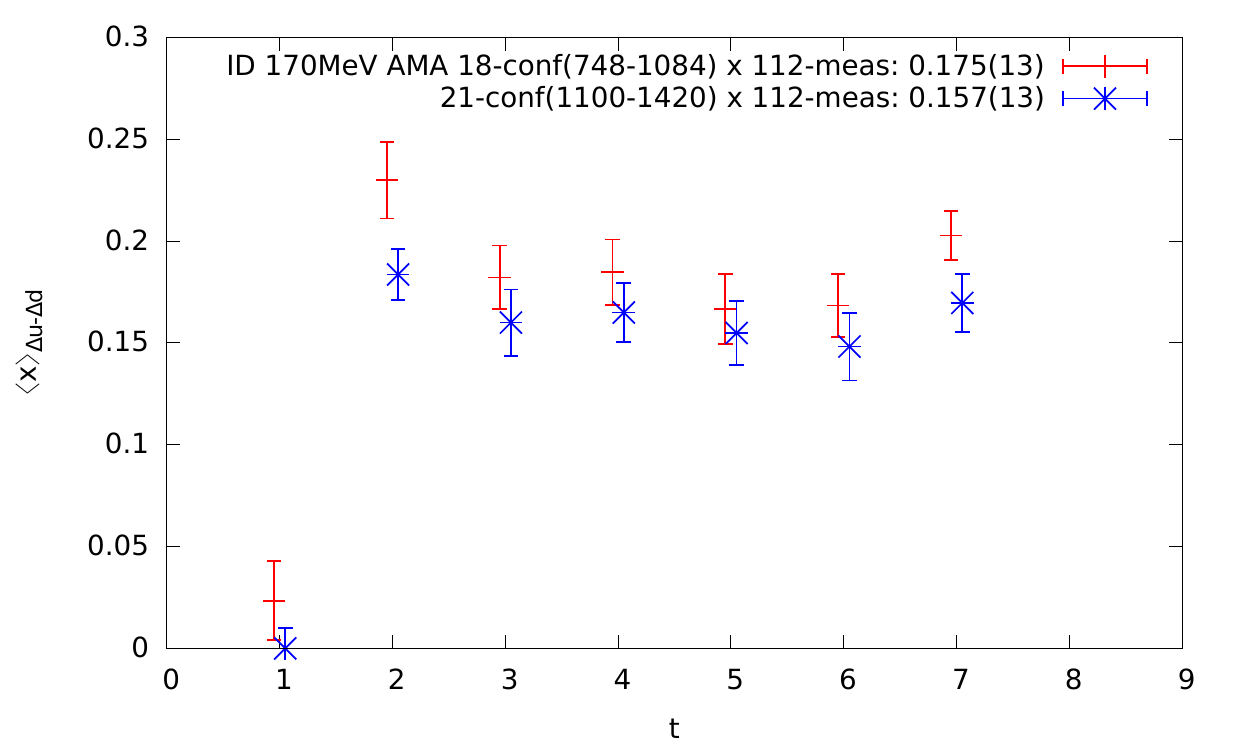}
\caption{\label{fig:transversity}
Top left: Isovector transversity, \(\langle 1 \rangle_{\delta u - \delta d}\), from the 170-MeV ensemble.
It may show some hint of similarly inefficient sampling as seen in the axial charge, but definitely less drastic, and more like other observables that do not show such sign at all, such as the isovector vector charge, \(g_V\), top right,  quark momentum, \(\langle x \rangle_{u-d}\),  and helicity, \(\langle x \rangle_{\Delta u-\Delta d}\), lower left and right.
}
\end{figure}

This year I reported our exploration for the cause that gives rise to this axial-charge systematics:
First, I now have results for an additional observable, the isovector transversity, \(\langle 1 \rangle_{\delta u - \delta d}\).
This observable, obtained from an insertion of \(\gamma_5 \sigma_{\mu\nu}\), differs from the axial charge, with insertion of \(\gamma_5\gamma_\mu\), by only one extra \(\gamma\)-matrix factor.
It may show similar signs of inefficient sampling as seen in the axial charge, but if at all definitely less drastic (see Fig.\ \ref{fig:transversity},)
and much more like other observables that do not show such sign at all, such as the isovector quark momentum fraction, \(\langle x \rangle_{u-d}\), on the right.
I also expanded blocked jackknife analyses to block sizes of 3 and 4 (see Table \ref{tab:blockJK}):
\begin{table}
\begin{center}
\begin{tabular}{cllll}
\hline
\multicolumn{5}{c}{Blocked jackknife analysis}\\
 & \multicolumn{4}{c}{block size}\\
 & \multicolumn{1}{c}{1}&\multicolumn{1}{c}{2}&\multicolumn{1}{c}{3}&\multicolumn{1}{c}{4}\\
\hline
\(g_V\) & 1.447(8) & 1.447(6) & \multicolumn{1}{c}{-} & \multicolumn{1}{c}{-} \\
\(g_A\) & 1.66(6) & 1.66(7) & 1.71(8) & 1.65(4) \\
\(g_A/g_V\) & 1.15(4) & 1.15(5) & 1.15(6) & 1.14(3) \\
\(\langle x \rangle_{u-d}\) & 0.146(7) & 0.146(8) & 0.146(8) & \multicolumn{1}{c}{-} \\
\(\langle x \rangle_{\Delta u - \Delta d}\) & 0.165(9) & 0.165(11) & 0.165(10) & \multicolumn{1}{c}{-} \\
\(\langle x \rangle_{u-d}/\langle x \rangle_{\Delta u-\Delta d}\) & 0.86(5) & 0.86(4) & \multicolumn{1}{c}{-} & \multicolumn{1}{c}{-} \\
\(\langle 1 \rangle_{\delta u - \delta d}\) & 1.42(4) & 1.42(6) & 1.42(6) & 1.41(3) \\
\hline
\end{tabular}
\end{center}
\caption{\label{tab:blockJK}
The error does not grow with block size except for the axial charge and transversity.}
\end{table}
the statistical error of the axial charge keeps growing to at least beyond block size of 3 while that for transversity stops growing earlier.
In other words these two observables behave similarly, resulting in the systematics that is much stronger in the axial charge.

\begin{figure}[b]
\begin{center}
\includegraphics[width=0.48\textwidth,clip]{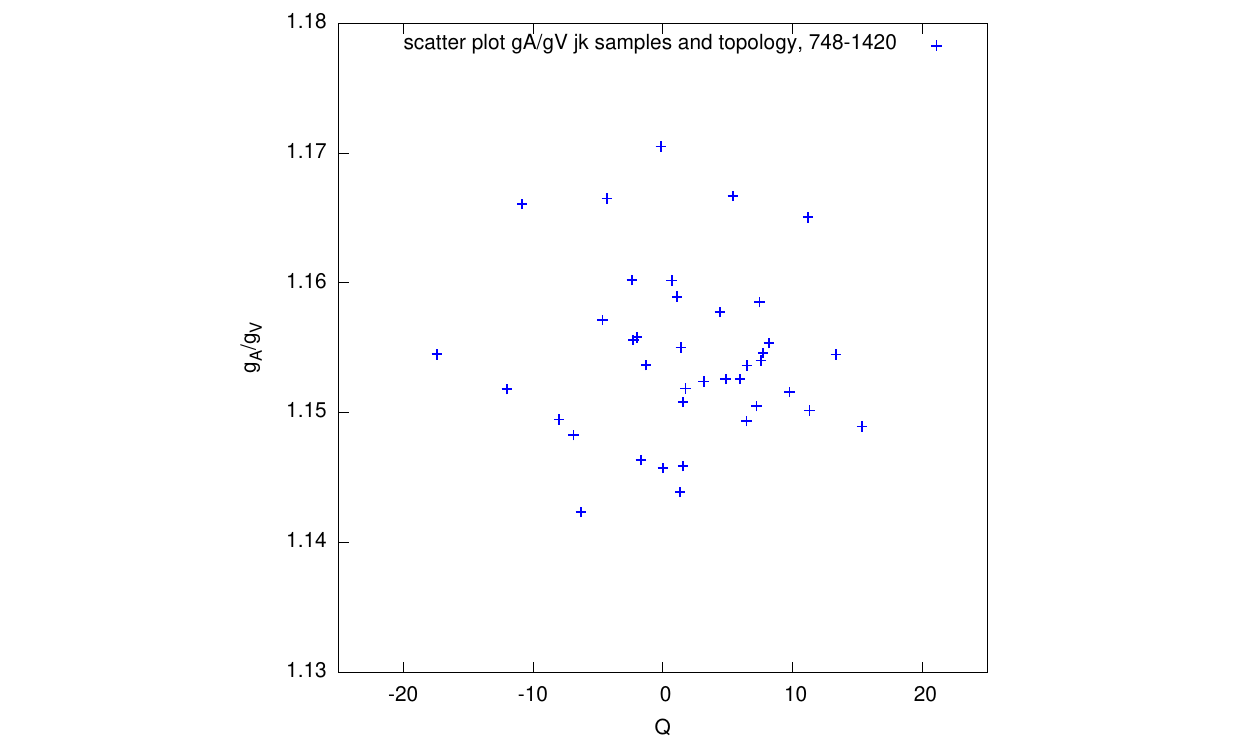}
\includegraphics[width=0.48\textwidth,clip]{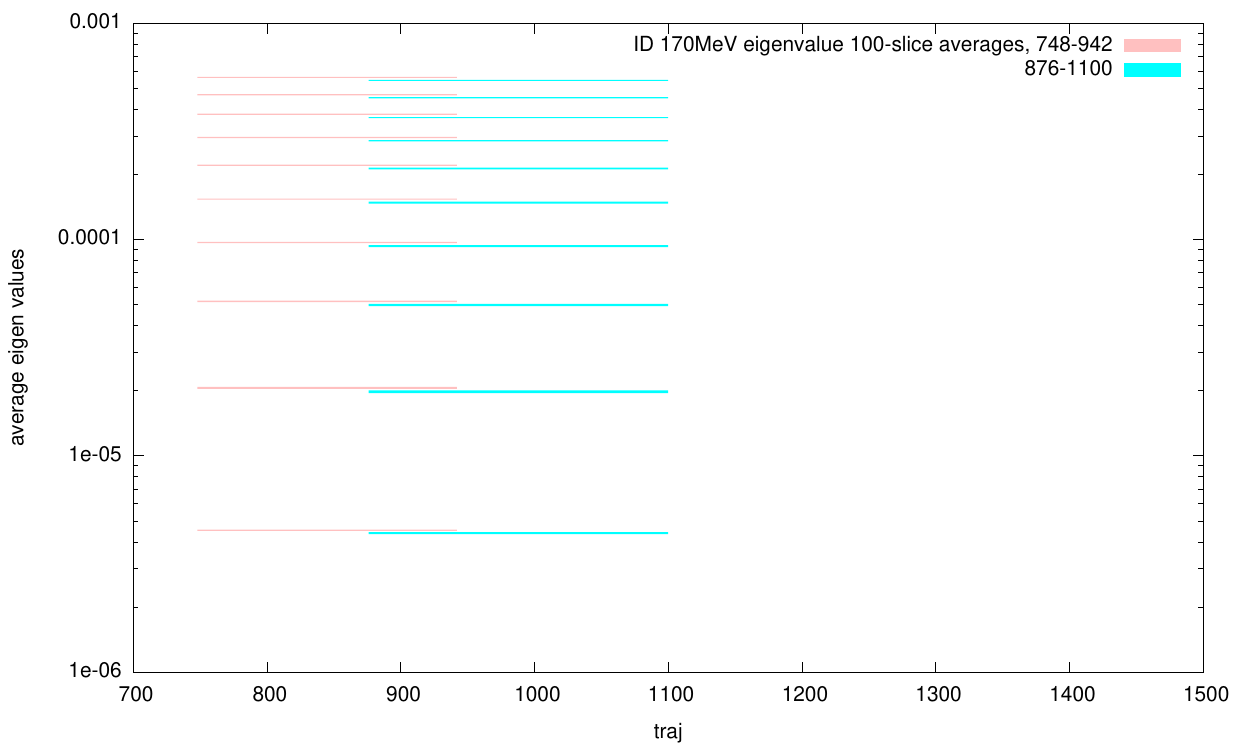}
\caption{\label{fig:TopDef}
Left: scatter plot of \(g_A/g_V\) jackknife samples against gauge topological charge:
No correlation is seen.
Right: Deflation eigenmode statistics for the two quarters where data are available:
there is no difference in the lowest 100 modes averaged.
Some insignificant difference emerges in higher modes.
}
\end{center}
\end{figure}

If an observable appears long-range autocorrelated, it would be interesting to look at its correlation with the topology of the gauge configurations.
We explored this possibly by plotting jackknife samples against topological charge (see left pane in Fig.\ \ref{fig:TopDef}),
and did not find correlation.

We can also look at if our low-mode deflation affected this, though the available information is limited to about half of the configurations of what we are presenting from the 170-MeV ensemble (see right pane of Fig.\ \ref{fig:TopDef}.)
Albeit with this limitation we do not find any correlation either:
the lowest 100 eigenmodes average do not differ between the two halves.
Some difference emerges as we go to higher eigenmodes but do not appear significant.

\begin{figure}
\begin{center}
\includegraphics[width=0.59\textwidth,clip]{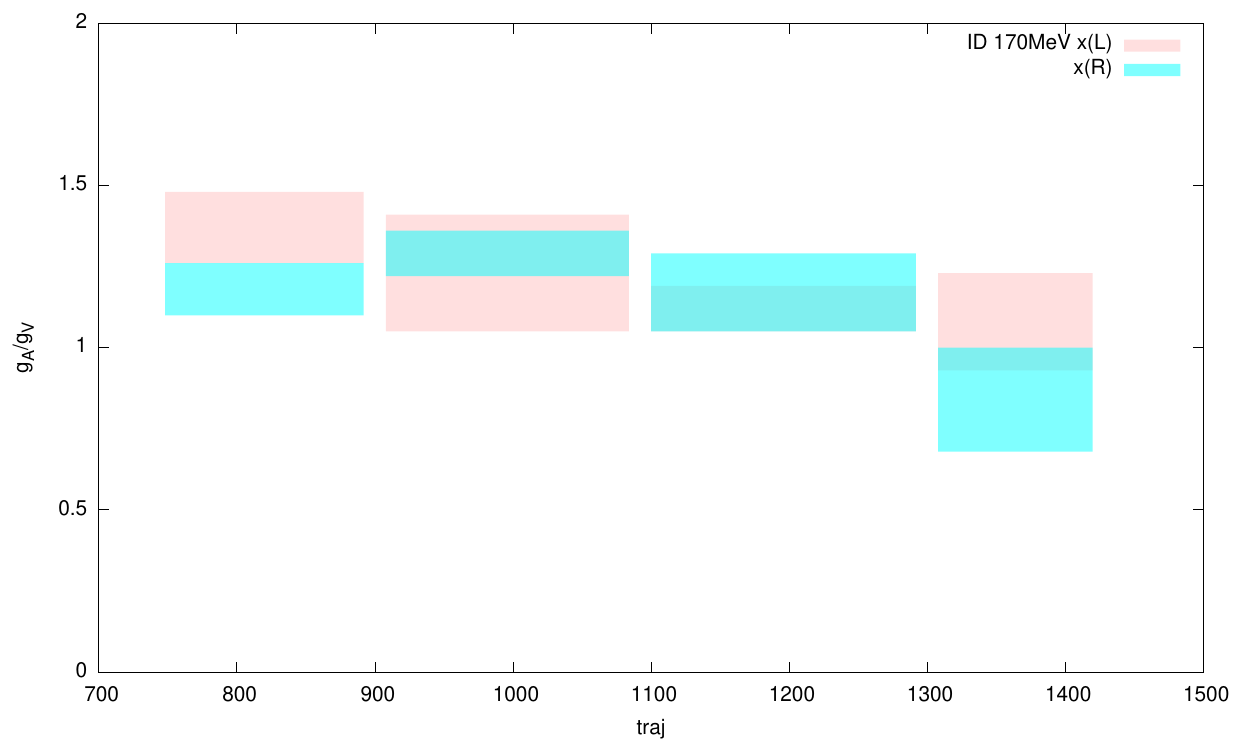}\\
\includegraphics[width=0.59\textwidth,clip]{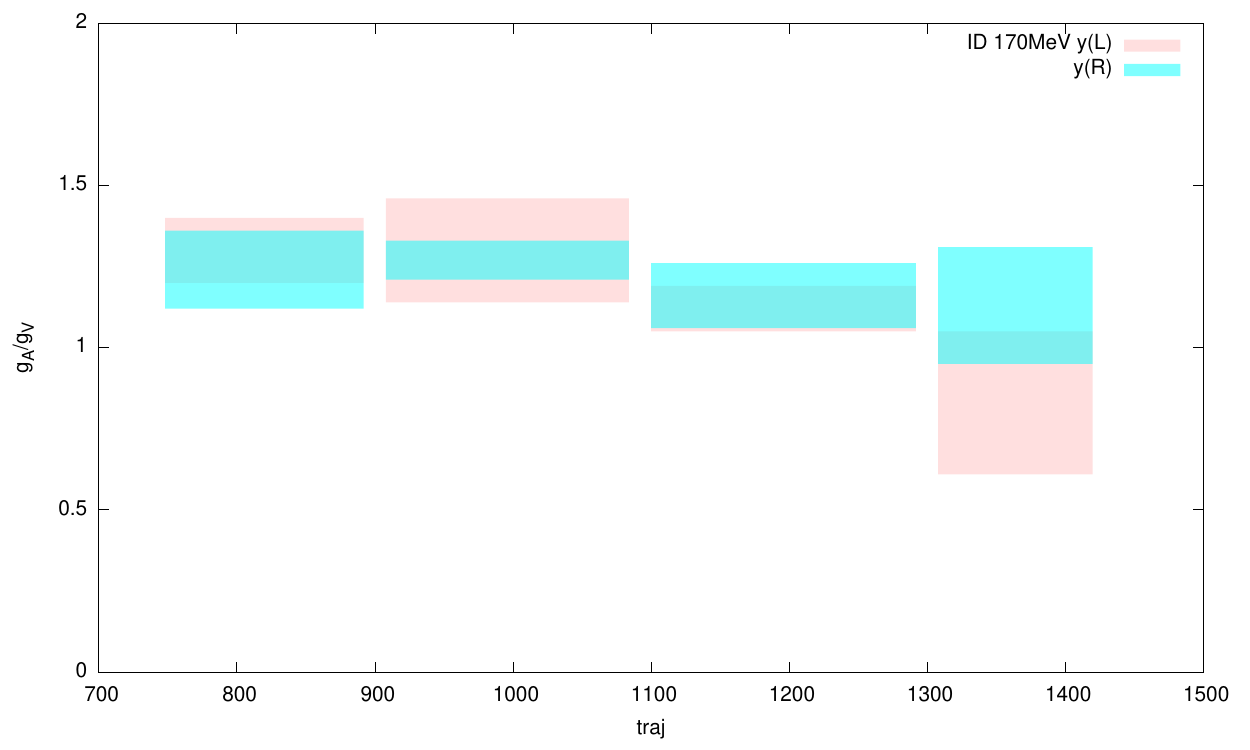}\\
\includegraphics[width=0.59\textwidth,clip]{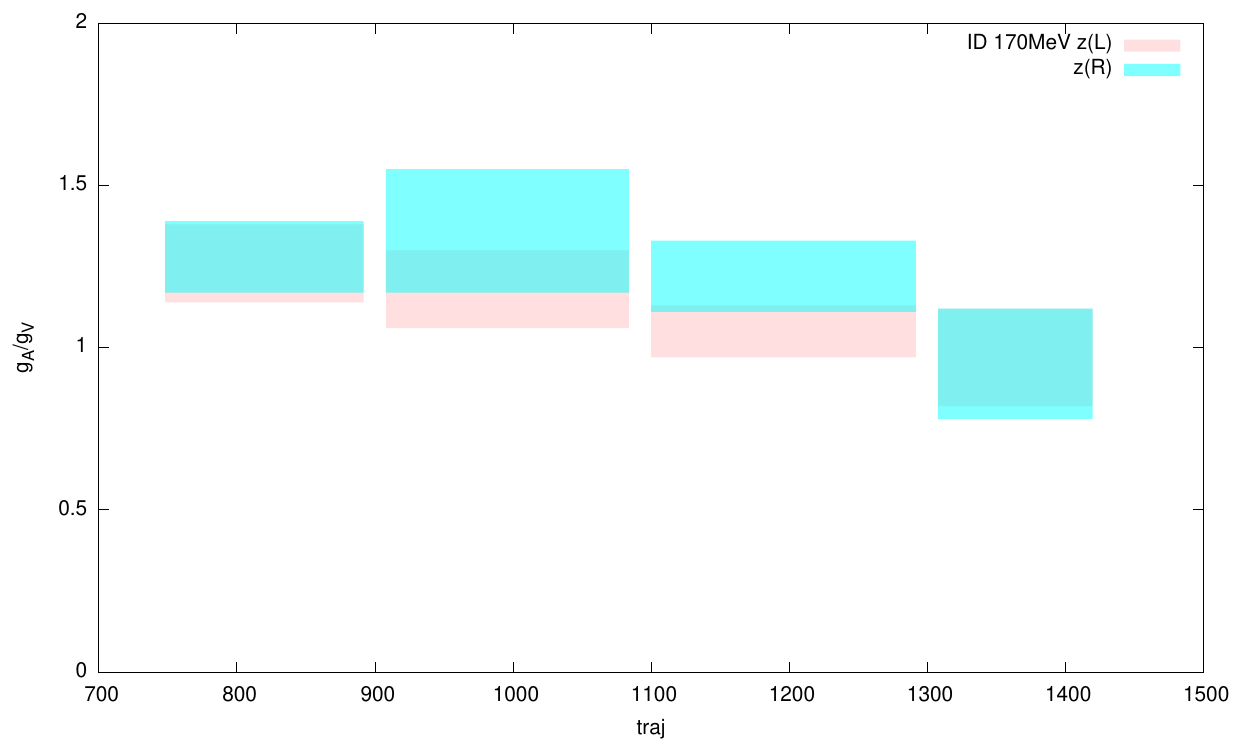}\caption{\label{fig:piongeometry}
Evolution of the ratio, \(g_A/g_V\), along the course of molecular dynamics time, divided into two spatial halves in \(x-\) (top), \(y-\) (middle) and \(z-\) (bottom) directions.
The calculation appears to fluctuate spatially. 
Spin polarization is along the \(z-\)axis.
}
\end{center}
\end{figure}

On the other hand, as was also reported last year, similar long-range autocorrelation was seen in the 330-MeV ensemble \cite{Ohta:2013qda} that is at the second smallest \(m_\pi L\), but not in the lighter 250-MeV nor the heaviest 420-MeV ensembles with larger \(m_\pi L\), hinting that the systematics may arise from the finite-size effect.
It may be also instructive to remember earlier phenomenological analyses such as by the MIT bag model that estimates \(g_A/g_V = 1.09\) without pion \cite{Chodos:1974pn}, and another by the Skyrmion model that gives only conditionally convergent result of 0.61, that is strongly dependent on pion geometry \cite{Adkins:1983ya}.
To explore such spatial dependence arising from pion geometry, we divided the AMA samples into two spatial halves such as \(0\le x < L/2\) and\(L/2 \le x < L\) for each of the three spatial directions in order to check if there is any uneven spatial distribution (see Fig. \ref{fig:piongeometry}.)
We found the calculation fluctuates in space.
Larger spatial volume would stabilize the calculation better.

\section{Summary}

We explored what causes about 10-\% deficit in lattice-QCD calculations of nucleon isovector axial and vector charge ratio, \(g_A/g_V\), in comparison with the experiment of 1.2723(23).
As we reported last year at Lattice 2013, an unusually long-range autocorrelation was seen in our lightest, 170-MeV, 2+1-flavor dynamical DWF ensemble.
It is also present in the 330-MeV ensemble with the second smallest \(m_\pi L\), but is not present in the other two, 250-MeV and 420-MeV ensembles with larger \(m_\pi L\).
No other isovector observable shows such a problem, except perhaps transversity where the effect is weak at worst.
No correlation is seen with gauge-field topology nor low-mode deflation.
In contrast the axial-charge calculation appears to fluctuate spatially along the course of molecular dynamics evolution.

I thank T.~Blum, T.~Izubuchi, C.~Jung, M.~Lin and E.~Shintani for close collaboration, and members of the RBC and UKQCD collaborations who contributed the four DWF ensembles.
The ensembles were generated using four QCDOC computers of Columbia University, Ediburgh University, RIKEN-BNL Research Center (RBRC) and  USQCD collaboration at Brookhaven National Laboratory, and a Bluegene/P computer of Argonne Leadership Class Facility (ALCF) of Argonne National Laboratory provided under the INCITE Program of US DOE.
Calculations of nucleon observables were done using RIKEN Integrated Cluster of Clusters (RICC) at RIKEN, Wako, and various Teragrid and XSEDE clusters of US NSF.

\end{document}